\documentclass[12pt,preprint]{aastex}

\newcommand{\be}{\begin{equation}}
\newcommand{\ee}{\end{equation}}
\newcommand{\bea}{\begin{eqnarray}}
\newcommand{\eea}{\end{eqnarray}}
\newcommand{\go}{\mathrel{\raise.3ex\hbox{$>$}\mkern-14mu\lower0.6ex\hbox{$\sim$}}}
\newcommand{\lo}{\mathrel{\raise.3ex\hbox{$<$}\mkern-14mu\lower0.6ex\hbox{$\sim$}}}

\shorttitle{Eccentricities of Planets from Photometry}
\shortauthors{Ford et al.~}

\begin{document}
\title{Characterizing the Orbital Eccentricities of Transiting Extrasolar Planets with Photometric Observations}

\author{Eric B.\ Ford\altaffilmark{1,2,3}, Samuel N.\ Quinn\altaffilmark{2}, Dimitri Veras\altaffilmark{3,4,5}}

\altaffiltext{1}{Hubble Fellow}
\altaffiltext{2}{Harvard-Smithsonian Center for Astrophysics, Mail Stop 51, 60 Garden Street, Cambridge, MA 02138, USA} 
\altaffiltext{3}{Department of Astronomy, University of Florida, 211 Bryant Space Science Center, P.O. Box 112055, Gainesville, FL 32611-2055, USA}
\altaffiltext{4}{JILA, Campus Box 440, University of Colorado, Boulder, CO 80309, USA}
\altaffiltext{5}{Department of Astrophysical and Planetary Sciences, University of Colorado, Boulder, CO 80309, USA}

\begin{abstract}
The discovery of over 200 extrasolar planets with the radial velocity
(RV) technique has revealed that many giant planets have large
eccentricities, in striking contrast with most
of the planets in the solar system and prior theories of planet
formation.  The realization that many giant planets have large
eccentricities raises a fundamental question: ``Do terrestrial-size
planets of other stars typically have significantly eccentric orbits
or nearly circular orbits like the Earth?''  
Here, we demonstrate that photometric observations of transiting
planets could be used to characterize the orbital eccentricities for
individual transiting planets, as well the eccentricity distribution
for various populations of transiting planets (e.g., those with a
certain range of orbital periods or physical sizes).  Such
characterizations can provide valuable constraints on theories for the
excitation of eccentricities and tidal dissipation.  We outline the
future prospects of the technique given the exciting prospects for
future transit searches, such as those to be carried out by the CoRoT
and Kepler missions.
\end{abstract}
\keywords{planetary systems --- planetary systems: formation --- planets 
and satellites: general  --- techniques: photometric --- methods: statistical
 --- celestial mechanics}

\section{Introduction}
\label{sec_intro}

Theorists have proposed numerous mechanisms that could excite orbital
eccentricities.  Some of these mechanisms are expected to affect all
planets independent of their mass (e.g., perturbations by binary
companions, passing stars, or stellar jets; e.g., Holman et al.\ 1997;
Laughlin \& Adams 1998; Ford et al.\ 2000; Zakamska \& Tremaine 2004;
Namouni 2007), while the efficiency of other mechanisms would vary
with planet mass (e.g., planet-disk or planet-planet interactions;
e.g., Artymowicz 1992; Goldreich \& Sari 2003; Chatterjee et al.\
2007).  If the mechanism(s) exciting eccentricities of the known giant
planets also affect terrestrial planets, then Earth-mass planets on
nearly circular orbits could be quite rare.  On the other hand, if
large eccentricities are common only in systems with massive giant
planets and/or very massive disks, then there may be an abundance of
planetary systems with terrestrial planets on low eccentricity orbits
(Beer et al.\ 2004).  Thus, understanding the eccentricity
distribution of terrestrial planets could provide empirical
constraints for planet formation theories (e.g., Ford \& Rasio 2007)
and shed light on the processes that determined the eccentricity
evolution in our solar system.  Since the discovery of transiting
giant planets with eccentric orbits, authors have begun to consider
the implications of eccentricities for transiting planets (e.g.,
Barnes 2007; Burke 2008).

The CoRoT and Kepler space missions are expected to detect many
transiting planets and measure their sizes and orbital periods,
including some in or near the ``habitable zone'' (e.g., Kasting et al.\
1993).  The Kepler mission aims to determine
the frequency of Earth-like planets, and study how the frequency and
properties of planets correlates with the properties of their host
stars (Basri et al.\ 2005).  
Since a significant eccentricity would cause the stellar flux incident
on the planet's surface to vary, a planet's eccentricity affects its
climate (i.e., equilibrium temperature, amplitude of seasonal
variability) and potentially its habitability (Williams et al.\ 2002;
Gaidos \& Williams 2004).  Our method could be applied to these
planets to determine the frequency of terrestrial planets that could
also have Earth-like climates, and thus influence the design of future
space missions that will attempt to detect and characterize nearly
Earth-like planets (e.g., Space Interferometry Mission-PlanetQuest)
and search them for signs of life (e.g., Terrestrial Planet Finder).

In \S\ref{SecDuration}, we show how the duration of a transit is
affected by a planet's orbital eccentricity.  We outline how to
interpret the transit duration for transiting planets with both low
signal-to-noise light curves (\S\ref{SecLoSN}) and high
signal-to-noise (\S\ref{SecHiSN}).  We compare the magnitude of the
effect on the transit duration to the expected precision of
eccentricity constraints based on Kepler photometric data
(\S\ref{SecStatErr}) and also the typical accuracy of stellar
parameters (\S\ref{SecStarErr}).
We demonstrate that Kepler observations could be used to characterize
the orbital eccentricities of terrestrial planets.  We describe
statistical approaches for analyzing the distribution of transit
durations of a population of transiting planets in \S\ref{SecStats}.
In \S\ref{SecRV}, we discuss the role of radial velocity observations
for constraining eccentricities of Earth-like transiting planets.  In
\S\ref{SecDiscuss}, we conclude with a discussion of how the results
could contribute to understanding the formation and evolution of
terrestrial planets and address fundamental questions, such as ``What
is the frequency of terrestrial planets that have Earth-like
eccentricities?'' and ``What is the frequency of terrestrial planets
that pass through the habitable zone?''

\section{Light Curve of an Eccentric Transiting Planet}
\label{SecLightCurve}

\label{SecDuration}
\begin{figure}
\plotone{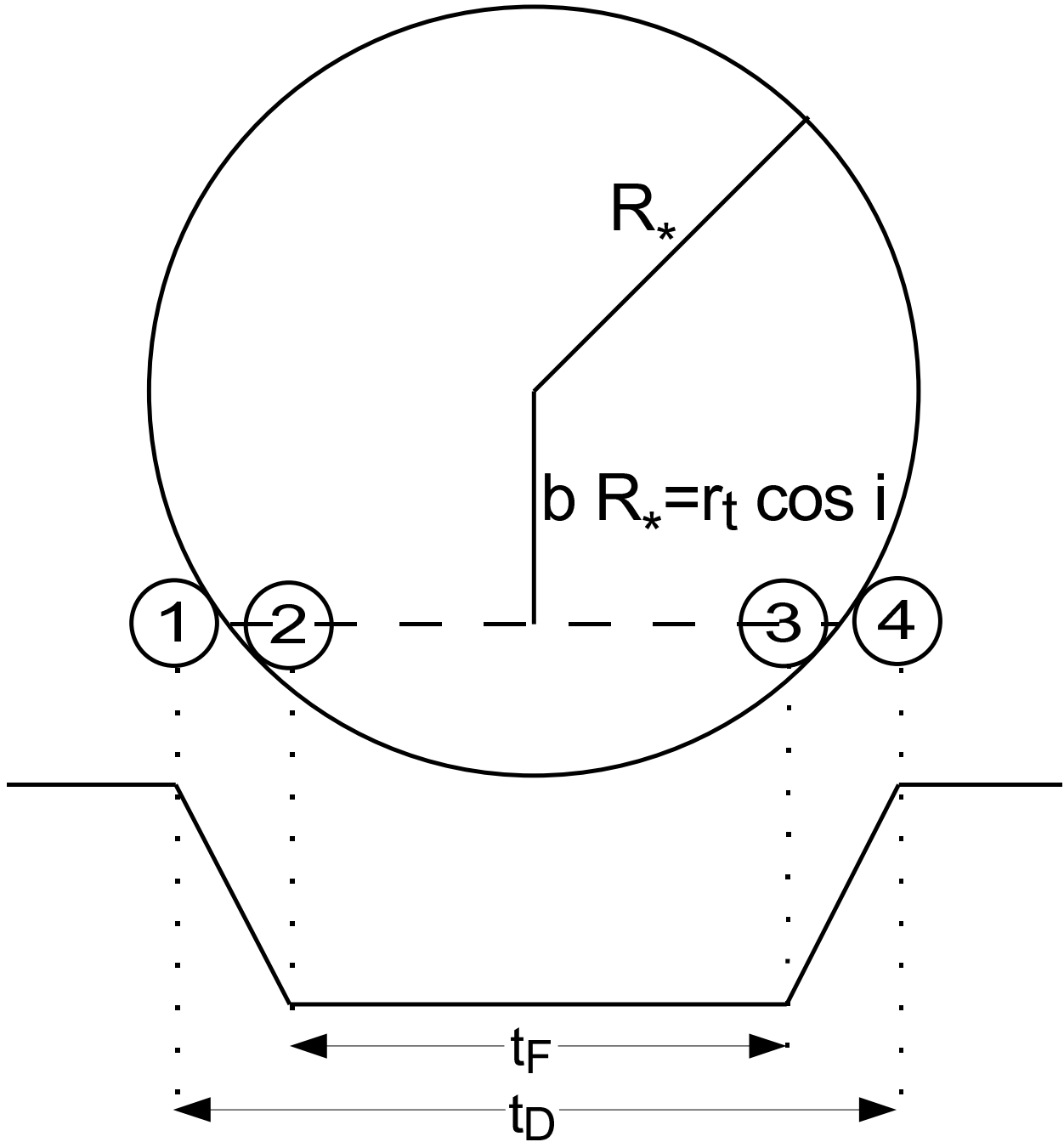} 
\caption{Geometry of Transit: Here we illustrate the
path (dashed line) of a planet (small circles) as it transits a star
(large circle).  The vertical dotted lines connect the planet to the
schematic light curve (solid curve) at the time of each of the four
points of contact.  Measuring the times of all four points of contact
allows for a measurement of the impact parameter ($b$) and
significantly increases the precision of the constraint on the
planet's eccentricity.  If only the total transit duration ($t_D$) is
measured, then the eccentricity distribution can still be constrained
for a larger population of transiting planets.
\label{FigGeo}}
\end{figure}

\begin{figure}
\plotone{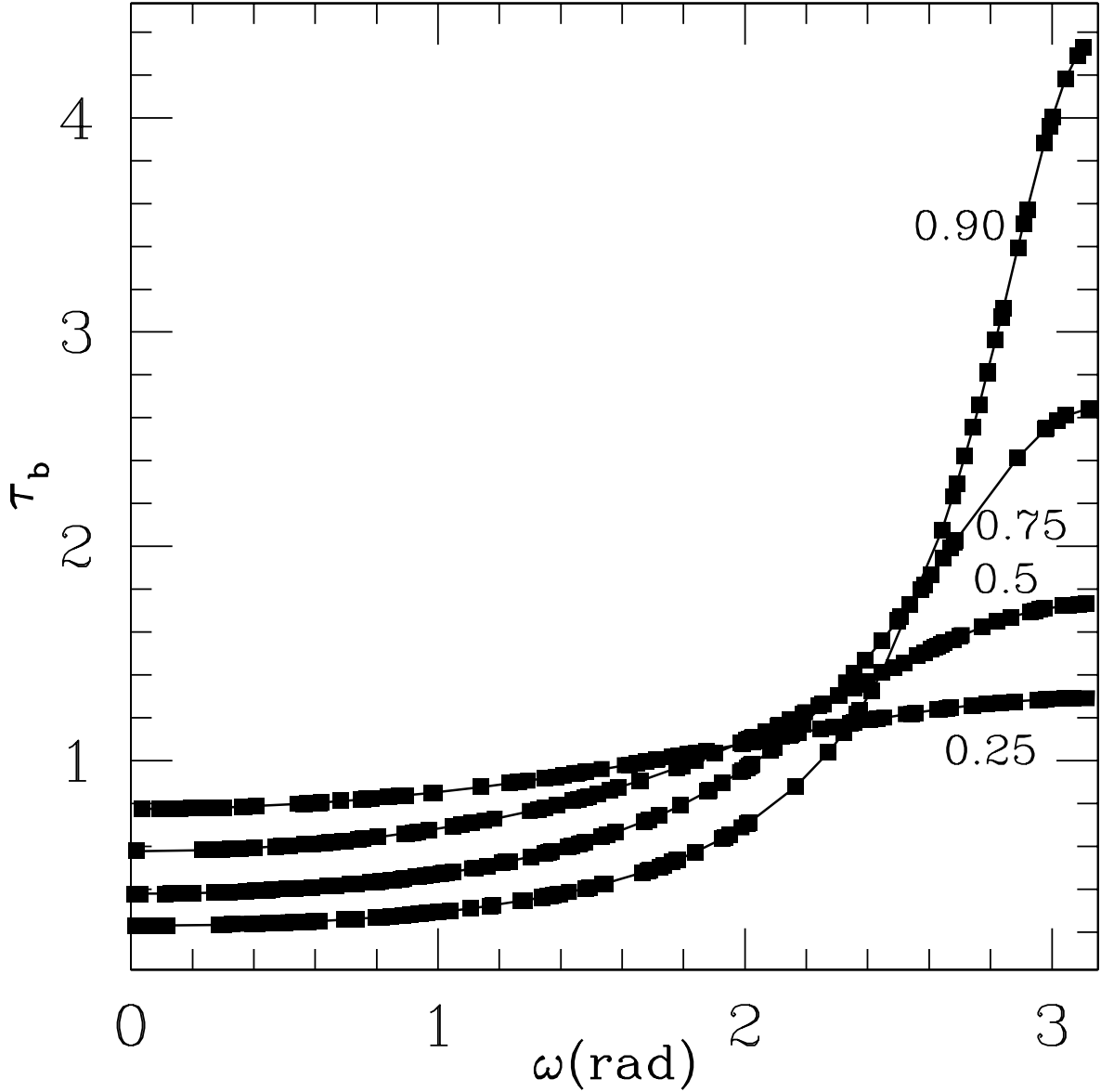} 
\caption{Transit Duration as a function of eccentricity ($e$) and argument of periastron ($\omega$):
The vertical axis shows $\tau_b$, the ratio of the actual transit
duration to the transit duration for a similar planet on a circular
orbit (with the same sizes, orbital period and impact parameter). The
curves show the analytic approximation for the transit duration (Eqn.\
1) for four values of eccentricity, 0.25, 0.5, 0.75, 0.9.  The points
are the exact durations for a planet on a Keplerian orbit.  For a
planet on a circular orbit, $\tau_b=1$, but for eccentric planets
$\tau_b$ can be larger (for planets that transit near apocenter) or
smaller (for planets that transit near pericenter).  Thus, a
measurement of the transit duration can be used to constrain a
combination of $e$ and $\omega$ (measured from the direction of the
observer), as well as to place a lower limit on the
eccentricity.
\label{fig_tauvsomega}   }
\end{figure}

The total transit duration is defined as the time interval between the
first and fourth points of contact (Fig. \ref{FigGeo}).  In the
approximation that the mean planet-star separation ($a$) is much
greater than both the stellar radius ($R_{\star}$) and the planetary
radius ($R_p$), the transit duration is much less than the orbital
period ($P$), and thus the planet-star separation during the transit
($d_t$) is nearly constant.  Using these approximations, the total
transit duration ($t_D$) is given by
\be 
\frac{t_D}{P} \simeq \frac{R_{\star}}{\pi a\sqrt{1-e^2}}
\sqrt{\left(1+r\right)^2-b^2} \left(\frac{d_t}{a}\right), 
\ee 
where $e$ is
the orbital eccentricity, $r\equiv R_p/R_{\star}$ is the ratio of
the planet radius to stellar radius, $b\equiv d_t \cos i / R_{\star}$
is the impact parameter, and $i$ is the orbital inclination
measured relative to the plane of the sky (Tingley \& Sackett 2005).
The planet-star separation at the time of transit is given by $d_t = a
( 1 - e \cos E_t ) =a (1-e^2)/(1+e \cos T_t)=a (1-e^2)/(1+e \cos
\omega)$, where $T_t$ and $E_t$ are the true and eccentric anomalies
at the time of transit, and $\omega$ is the argument of periastron
measured relative to the line of sight.  (Note that this differs from
the convention for radial velocity determinations.)  Thus, the ratio
of the actual transit duration to the transit duration for the same
planet on a circular orbit (assuming other transit parameters such as
size and impact parameter are held fixed) is $(1+e
\cos\omega)/\sqrt(1-e^2)$.  We have
performed numerical integrations to verify that this approximation is
accurate to better than $\simeq0.1\%$, for eccentricities as high as
0.95 (Fig.\ \ref{fig_tauvsomega}).

\begin{figure}
\plotone{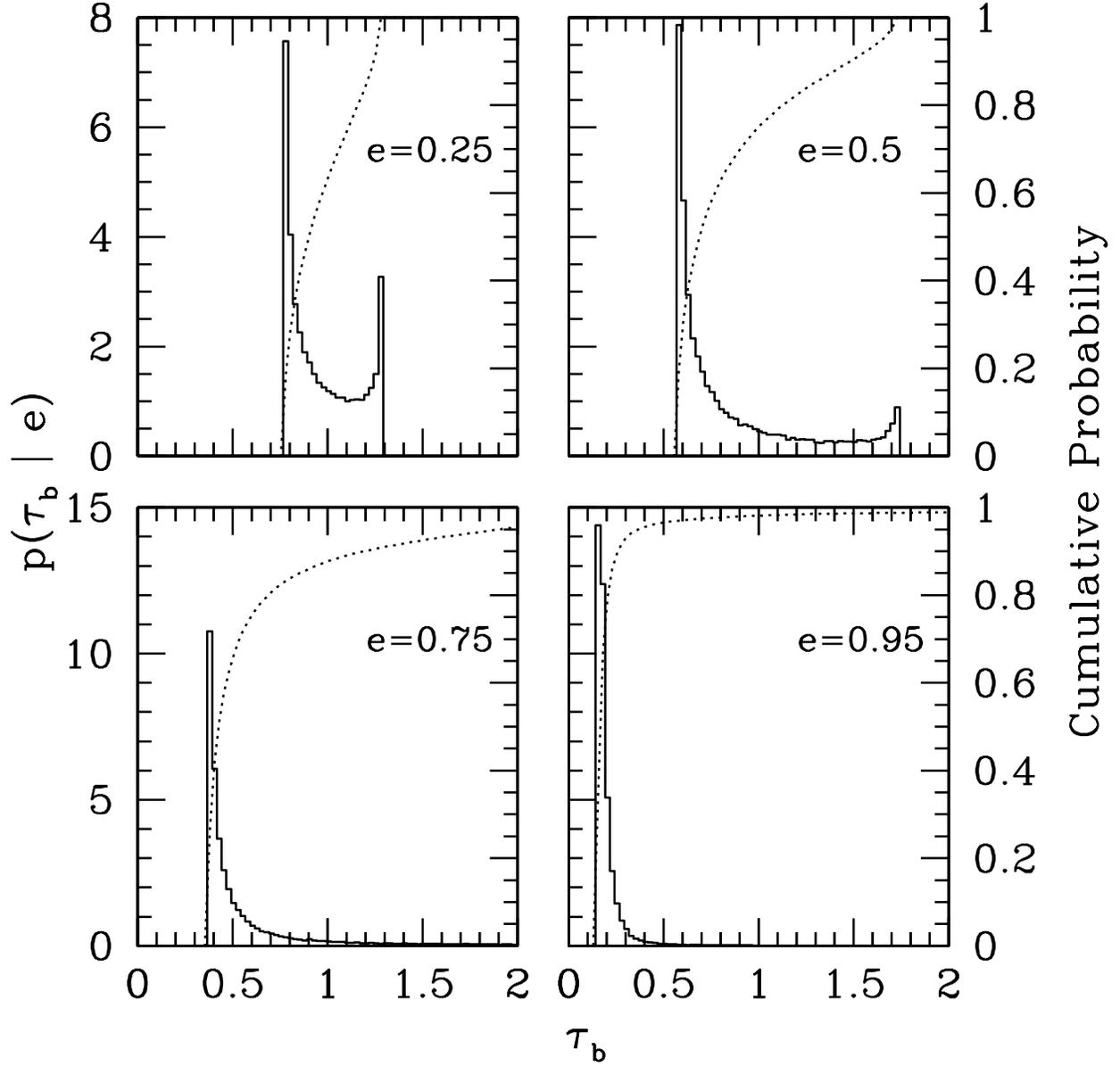} 
\caption{
Probability Distribution for Transit Duration at a given eccentricity ($e$):
Here we show the probability distribution (solid histogram) for
$\tau_b$ for an ensemble of planets with a single fixed eccentricity,
assuming a uniform distribution of the argument of pericenter and an
isotropic distribution of inclinations.  The dotted curves are the
cumulative probability distributions.  For large eccentricities, there
is a small fraction of very long transits ($\tau_b>2$ off the scale).
\label{FigProbTau} }
\end{figure}

We define the variable
%
$
\tau_b \equiv d_t/(a\sqrt{1-e^2})
$
%
to be the ratio of the transit duration of one planet to the transit duration for an
identical planet with the same orbital period and impact parameter,
but on a circular orbit.
In general, transiting planets could have significant eccentricities,
in which case $\tau_b$ will deviate from unity (e.g., $\tau_b\simeq1.8$ for HAT-P2b; Bakos et al.\
2007).  For planets on low eccentricity orbits,
$\left|\tau_b-1\right|\simeq O(e)$.  For large eccentricities, $\tau_b$
can range from zero to several (Fig.\ \ref{fig_tauvsomega}).
For the case of a central transit of an Earth-like planet
orbiting a solar-mass star on a circular orbit,
\be
%
t_D \simeq 13 \mathrm{hours} \left(\frac{P}{\mathrm{yr}}\right)^{1/3} \left(\frac{R_{\star}}{R_\odot}\right) \left(\frac{M_{\odot}}{M_{\star}}\right)^{1/3} \left(1+\mu\right)^{-1/3} \left(\left(1+r\right)^2-b^2\right)^{1/2}  \tau_b ,
\ee
where $M_{\star}$ is the stellar mass and $\mu\equiv m_p/M_{\star}$ is
the planet-star mass ratio.  The planet-star mass ratio ($\mu\equiv
m_p/M_{\star}$) is typically negligible for planetary companions.  (If
it is large, then it could be measured by obtaining RV observations.)
By assuming that the pericenter direction is randomly oriented relative
to the line of sight, we can compute the probability distribution for
$\tau_b$ as a function of eccentricity (Fig.\ \ref{FigProbTau}).

\subsection{Analysis of Transiting Planet Light Curves}
\label{SecAnalysis}
Here we outline the basics of how photometric light curves can be used
to constrain the eccentricities of transiting planets.  We discuss two
limiting regimes: 1) low signal-to-noise (S/N) light curves that
provide no constraint on the impact parameter, and 2) high S/N light
curves that measure the impact parameter.

\subsubsection{Impact Parameter Not Measured (Low S/N)}
\label{SecLoSN}
For any transit detection, photometric observations alone directly
measure the orbital period ($P$), the total transit duration ($t_D$),
and the depth of the transit.  For the purposes of providing analytic
estimates, we neglect the effects of limb darkening, so that the
transit depth is given by the planet-star area ratio ($r^2$).  
For some faint stars and/or small planets, it may be difficult to
measure additional light curve parameters such as the ingress duration
and the impact parameter.  For these cases, the total transit
duration, $t_D$, can be compared to $t_{D,o}$, the transit
duration expected for the same planet with the same orbital period
($P$) and mean stellar density ($\rho_{\star}$), but assuming a
circular orbit and central transit ($b=0$).  Even for transiting
planets with low S/N light curves, photometric observations can measure the ratio
\begin{equation}
\tau_o \equiv \frac{t_D}{t_{D,o}} = \left(\frac{d_t}{a\sqrt{1-e^2}}\right)  \left(\frac{\sqrt{\left(1+r\right)^2-b^2}}{1+r}\right) \simeq \left(\frac{t_D}{13\mathrm{hours}}\right) \left(\frac{\mathrm{yr}}{P}\right)^{1/3} \left(\frac{\rho_{\star}}{\rho_{\odot}}\right)^{1/3}  \frac{\left(1+\mu\right)^{1/3}}{\left(1+r\right)}.
\label{EqnTauo}
\end{equation}
Since $\mu=m_p/M_{\star}\ll 1$ for planetary mass companions, a value
of $\tau_o$ significantly greater than unity can only arise for an
eccentric planet, but a value of $\tau_o$ less than unity could be due
to either a non-central transit ($b>0$) or a non-zero eccentricity
($e>0$).  Thus, it is not possible to measure the
eccentricity for an individual planet with only low S/N photometry.
However, it is still possible to characterize the distribution of
eccentricities for a population of planets based on the observed
distribution of $\tau_o$ (see \S\ref{SecStatsKs}).  Since the above analysis can be applied
to relatively faint stars with low S/N transit light curves, we expect
that Kepler will discover many planets that can be included in 
statistical analyses of $\tau_o$ (Basri et al.\ 2005).  
We note the actual analysis should consider the effects of limb
darkening, which cause the depth of a high-latitude transit relative
to be greater than a than an equatorial transit even for the same
star-planet area ratio.  While limb darkening precludes simple
analytic expressions, such effects can be incorporated into Monte
Carlo simulations.

In
\S\ref{SecStats}, we will show that Kepler observations could
constrain the {\em distribution of eccentricities} and reject
plausible eccentricity distributions for terrestrial planets.
The precision of the eccentricity constraints will depend on the
precision and accuracy of the measurements of $\tau_o$.  In
\S\ref{SecStatErr} we show that the the orbital period, transit depth,
and transit duration can be measured with sufficient precision that
the accuracy for the measured $\hat{\tau}_o$ will typically be limited
by the deviation of the inferred stellar density
($\hat{\rho}_{\star}$) from the actual stellar density
($\rho_{\star}$), i.e., $\hat{\tau}_o\simeq \tau_o
(\hat{\rho}_{\star}/\rho_{\star})^{1/3}$.  The mean stellar density
could be estimated using spectroscopy and stellar modeling
(\S\ref{SecStarErr}), resulting in a typical accuracy for
$\hat{\tau}_o$ of $\sim 5-15\%$ (e.g., Ford et al.\ 1999; Fischer \& Valenti 2005; Takeda et al.\ 2007).  The spread of transit
durations due to the unknown inclination will have a negligible
effect on the ability of low S/N transit light curves to constrain the
eccentricity distribution.

\subsubsection{Impact Parameter Measured (High S/N)}
\label{SecHiSN}
For high-quality light curves, photometric observations can measure the time of each point of contact and determine both the
total transit duration ($t_D$) and $t_F$ (Fig.\ \ref{FigGeo}).  If we
neglect limb darkening, then 
%
%
the impact parameter can be determined by the photometric observables,
%
$
b^2 = \left[\left(1-r\right)^2-\gamma^2\left(1+r\right)^2\right]/\left(1-\gamma^2\right),
$
%
where we define $\gamma^2\equiv t_F^2/t_D^2$ (Seager \& Mallen-Ornelas
2003).  

\begin{figure}
\plotone{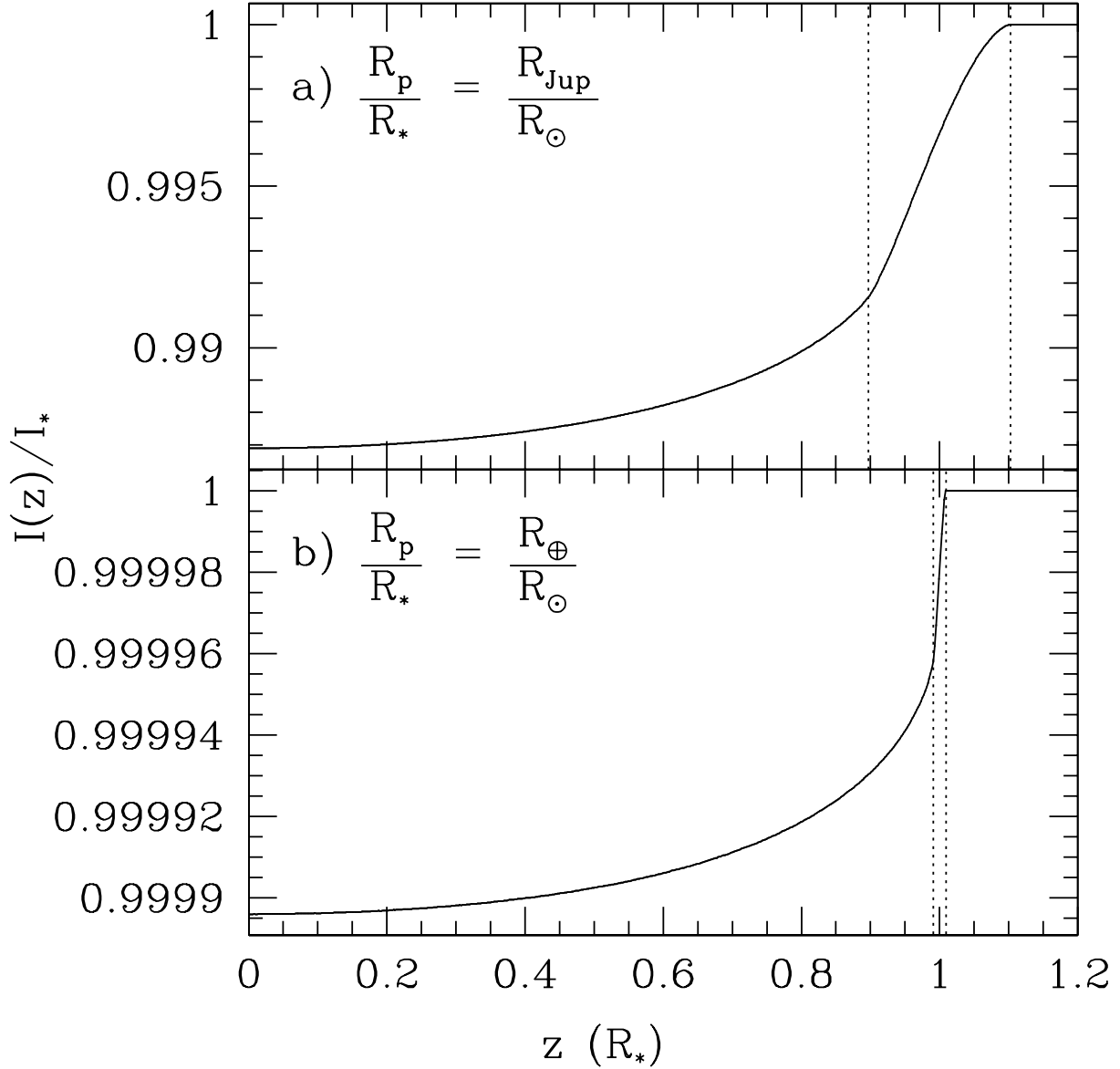} 
\caption{Transit light curves illustrating effects of
limb-darkening.
Here we show the observed intensity ($I(z)$; normalized to total
stellar flux, $I_{\star}$) as a function of $z$, the projected
separation from the center of the star (measured in stellar radii).
The vertical dotted lines indicate the points of contact.  For
this illustration, we adopt a
quadratic limb-darkening model (Mandel \& Agol 2002) and limb
darkening coefficients of $\gamma_1=0.4382$ and $\gamma_2=0.2924$
based on ATLAS models for solar-like star in V band (Claret 2000). 
{\em Top:} For a Jupiter-sized planet, 
%
%
{\em Bottom:} For an Earth-sized planet.
%
%
\label{FigLimbDarkening}   }
\end{figure}

\begin{figure}
\plotone{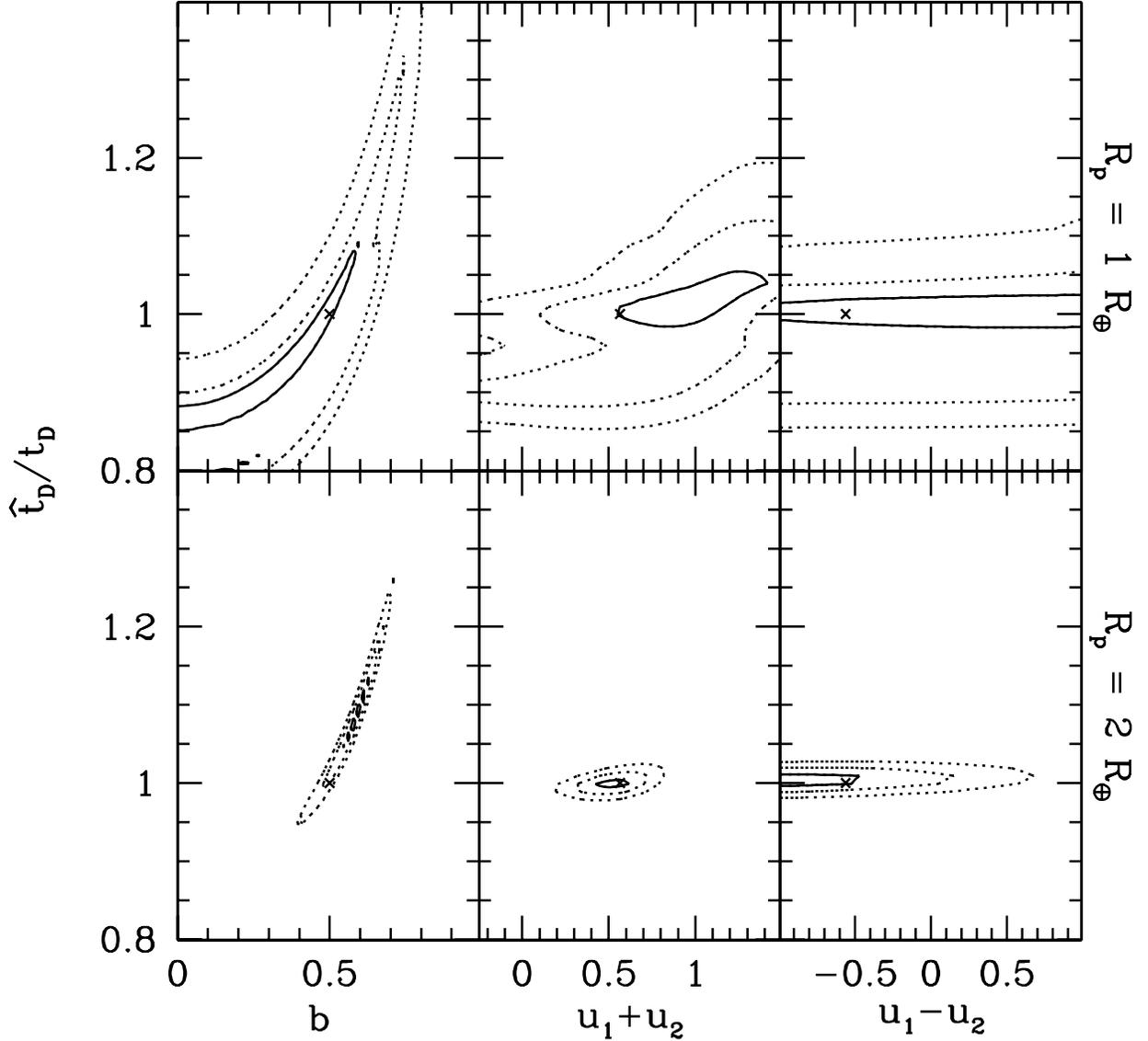} 
\caption{Correlations between transit duration and impact parameter or
limb darkening coefficients.  Here we shows contour of constant
$\Delta\chi^2$ = 1, 4, and 9 with $\hat{t}_D/t_D$ (the ratio of the
model transit duration to the actual transit duration) on the
$y$-axis.  The various collumns have $x$-axes of the impact parameter
($b$; left), the sum of the quadratic limb darkening coefficients
($u_1+u_2$; center), and the difference of the quadratic limb
darkening coefficients ($u_1-u_2$; right).  The top row shows a 1
$R_\oplus$ planet and the bottom row shows a $2 R_\oplus$ planet, both
assumed to be at 1 AU from a solar mass star.
\label{FigContours} }
\end{figure}

In practice, the transit shape will be affected by limb darkening
(Fig.\ \ref{FigLimbDarkening}) and we must fit for the limb darkening
parameters to determine the impact parameter (and planet-star radius
ratio).  Previous experience analyzing transit photometry of giant
planets has found that the limb darkening parameters can be accurately
estimated with stellar models and the uncertainty in limb darkening
parameters typically introduces relatively modest additional
uncertainties and have only small correlations with the other model
parameters of interest (e.g., Brown et al.\ 2001; Holman et al.\ 2006;
Burke et al.\ 2007; Knutson et al.\ 2007; Winn et al.\ 2007; Torres et
al.\ 2008).  
In order to verify that the uncertainty in limb darkening coefficients
will introduce only modest additional uncertainties, we have generated
simulated Kepler V-band light curves for terrestrial planets
transiting a $V=12$ magnitude solar-like star, including a quadratic
limb darkening model (Mandell \& Agol 2002; Claret 2000).  We then
calculate $\Delta\chi^2$, varying the transit time, transit duration,
the planet size, impact parameter, and quadratic limb darkening
coefficients (see Fig.\ \ref{FigContours}).  While these
are some modest correlations between the transit duration and limb
darkening parameters, these correlations are a much smaller effect
than the uncertainty due to the correlation between the transit
duration and the impact parameter.  Therefore, we conclude that the
uncertainty in impact parameter will be the limiting factor and we can
neglect the effects of limb darkening when estimating the power of
transit durations for constraining the eccentricities of transiting
planets (see \S\ref{SecStatsKs}).
%
%
Once the host star and transit time are known from Kepler
observations, a future transit can be observed using ground-based
observatories at multiple wavelengths.  By combining the inferred
stellar properties with limb darkening profiles from stellar models,
one can determine the impact parameter from even relatively low
precision data (e.g., Jha et al.\ 2000).
Infrared observations would be
particularly useful, as they minimize the effects of limb darkening
and aid in the measurement of the impact parameter.  Since the impact
parameter is the same at all wavelengths, ground-based transit light
curves could effectively eliminate the need to fit for this parameter
when analyzing the Kepler photometry.  Here, we assume that limb-darkening
parameters can be well constrained by some combination of stellar modeling
and external observations.  
%

For systems with high S/N photometry, the total transit duration,
$t_D$, can be compared to $t_{D,b}$, the transit duration expected for
the same planet and star with the same observed orbital period, impact
parameter ($b$), and mean stellar density, but
assuming a circular orbit. Then, 
Kepler can measure the ratio
\begin{equation}
\tau_b = \frac{t_D}{t_{D,b}} \simeq \left(\frac{t_D}{13\mathrm{hours}}\right) \left(\frac{\mathrm{yr}}{P}\right)^{1/3}  \left(\frac{1-\gamma^2}{r}\right)^{1/2} \left(\frac{\rho_{\star}}{\rho_{\odot}}\right)^{1/3}  \left(1+\mu\right)^{1/3}.
\label{EqnTauB}
\end{equation}
The fact that the planet is transiting the star places a constraint on a
combination of true anomaly and the direction of pericenter.  Therefore, 
the actual value of $\tau_b$ depends only on the eccentricity and
direction of pericenter, and $\tau_b$ can be measured using a combination
of transit photometry and stellar parameters (e.g.,  from spectroscopy and stellar modeling).
Using high S/N photometry, we can accurately measure the first
three terms on the right hand side of Eqn. \ref{EqnTauB}.  The last term containing the planet-star
mass ratio ($\mu$) is negligible for planetary
companions.
Therefore, the accuracy of the measured $\tau_b$ will typically be
limited by the uncertainty in the cube root of the mean stellar density
($\sigma_{\rho_{\star}^{1/3}}$).
We can detect a non-zero eccentricity for 
an individual planet when
$e>\sigma_{\rho_{\star}^{1/3}}/\rho_{\star}^{1/3}\simeq
0.05-0.15$  (\S\ref{SecStarErr}).  This is quite significant given the mean eccentricity 
of planets discovered by RV surveys is $\simeq 0.3$ (ignoring planets
with $P<10$ days that may have been influenced by tidal circularization; Butler et al.\ 2006).

\subsection{Expected Precision of Kepler Measurements}
\label{SecStatErr}
Here we present estimates of the timing precision possible with Kepler
observations, assuming a flux measurement precision of
$\sigma_{ph}\simeq$400ppm during a one minute integration on a V=12
star and uncertainties that scale with the square root of the photon
count (Basri et al.\ 2005).  For the sake of deriving approximate
analytic expressions, we assume uncorrelated Gaussian uncertainties
and ignore complications due to limb darkening in this paper

To determine the precision of eccentricity constraints, we must
estimate the precision of the constraints from transit photometry.
The fractional transit depth ($r^2$ in the absense of limb darkening)
can be measured with a precision, $\sigma_{r^2}\simeq \sigma_{ph} /
\sqrt{t_F \Lambda N_{tr}}$, where $\Lambda$ is the rate of photometric
measurements with precision $\sigma_{ph}$ and $N_{tr}\simeq 4
(\mathrm{yr}/P)$ is the number of transits observed during the four
year mission lifetime.  Note that in this approximation the precision
does not depend on the integration time, provided that it is
significantly less than the ingress/egress time.  In practice, an
increased rate of measurements is valuable for constraining any
brightness variations across the stellar disk (e.g., limb darkening,
star spots, plage).

The time of
each ingress/egress ($t_{in/out}$) can be measured with a precision
$\sigma_{t_{in/out}} \simeq \sigma_{ph} \sqrt{\Delta t_{in/out}/\Lambda}
\left(R_{\star}/R_p\right)^{2}$, where $\Delta t_{in/out}$ is the duration of ingress/egress (Ford \& Gaudi 2006).
Therefore, the sidereal orbital period can be measured with a
precision of $\sigma_P \simeq \sigma_{t_{in/out}}
/\sqrt{2\left(N_{tr}-1\right)}$ and the duration of the transit can be
measured with a precision of $\sigma_{t_D} = \sigma_{t_{in/out}}
\sqrt{2/N_{tr}}$.  Once there are at least two transits, the
uncertainty in $t_D$ will dominate.

The duration of ingress ($\Delta t_{in}$) or egress($\Delta t_{out}$) is given by
$\Delta t_{in/out}\equiv \epsilon t_D$, where $\epsilon
\left(1-\epsilon\right) =
r/\left[\left(1+r\right)^2-b^2\right]$.  For non-grazing
transits, $\epsilon\ll1$, so the ingress/egress duration is
%
%
%
%
\be
\Delta t_{in} \simeq \Delta t_{out} \simeq 7.1 \mathrm{minutes} \left(\frac{R_p}{R_\oplus}\right) \left(\frac{P}{\mathrm{yr}}\right)^{\frac{1}{3}} \left(\left(1+r\right)^2-b^2\right)^{\frac{-1}{2}} \left(\frac{M_{\odot}}{M_{\star}}\right)^{\frac{1}{3}} \left(1+\mu\right)^{\frac{-1}{3}} \tau_b.
\ee
Inserting fiducial values for a target star with an apparent magnitude of $V$, 
%
%
%
we expect a fractional precision for the mean transit duration of an individual planet to be 
\be
\frac{\sigma_{t_D}}{t_D} \simeq 0.013 \left(\frac{P}{\mathrm{yr}}\right)^{\frac{1}{3}} \left(\frac{R_p}{R_\oplus}\right)^{\frac{-3}{2}}  \left(\frac{R_{\star}}{R_\odot}\right)  \left(\frac{M_{\star}}{M_{\odot}}\right)^{\frac{1}{6}} \left(1+\mu\right)^{\frac{1}{6}} \tau_b^{\frac{-1}{2}} \left(\left(1+r\right)^2-b^2\right)^{\frac{-3}{4}} 10^{\left(V-12\right)/5}.
\ee
For a Jupiter-sized planet with a similar orbit and host star, the
expected fiducial fractional uncertainty for Kepler observations
decreases to $\simeq3\times10^{-4}$.  Ground-based
observatories have achieved a precision of
$\sigma_{t_D}/t_D\simeq1.5\%$ for short-period giant planets (e.g.,
Holman et al.\ 2007).

For modest eccentricities, Kepler's expected measurement
precision will place a lower limit on the uncertainty of a given
planet's eccentricity ($\sigma_e$) of order $\simeq \sigma_{t_D}/t_D$.
Even for Earth-sized planets around V=14 stars,
Kepler is expected to achieve the photometric precision necessary to
measure eccentricities as low as $\simeq 0.03$.  

For the sake of completeness, we also estimate one additional
timescale. For an eccentric orbit, there is a slight change in $d_t$
between ingress and egress that results in a difference between the
ingress and egress durations,
\be
%
\Delta t_{in}-\Delta t_{out} \simeq 4 \mathrm{sec} \left(\frac{P}{\mathrm{yr}}\right)^{-1/3} \left(\frac{R_p}{R_\oplus}\right) \left(\frac{R_{\star}}{R_\odot}\right) \left(\frac{M_{\odot}}{M_{\star}}\right)^{2/3} \left(\frac{\left(1+\mu\right)^{-2/3}  e \sin T_t}{\left(1-b^2\right)  \left(1+e\cos T_t\right) \left(1-e^2\right)^{3/2} }\right),
\ee
where $T_t$ is the true anomaly at the time of mid-transit and we have
neglected the second order effect of change in the impact parameter.
For a Jupiter-sized planet with a 4 day orbit around a solar-sized star,
the difference in ingress and egress durations increases to 3.4
minutes.  While Kepler and/or high-precision follow-up observations
might be able to measure such effects for a relatively small number of
bright stars, we expect other methods will typically provide more
powerful constraints on the eccentricity.

\subsection{Uncertainties in Stellar Mass and Radius}
\label{SecStarErr}
For many transiting planets photometric observations will be so
precise that uncertainties in the stellar properties will limit the
accuracy of $\hat{\tau}_o$ or $\hat{\tau}_b$ and hence the
eccentricity constraints (unless there are significant constraints
from RV observations or time of secondary transit; see \S\ref{SecRV}).
From the perspective of stellar modeling, the physical properties of a
star (e.g., radius, and density) are a function of at least three key
variables: stellar mass, composition, and age.  Hence stellar modeling
can only provide powerful constraints on the star's physical
properties if there are at least three observational constraints.  The
effective temperature ($T_{\rm eff}$) and metallicity ($[Fe/H]$) can
be accurately derived from a single high-precision spectroscopic
observation.  When a precise parallax ($\pi$) is available, the
stellar luminosity ($L_{\star}$) can be calculated from the apparent
magnitude (V; and a bolometric correction).  The three constraints
($T_{\rm eff}$, $[Fe/H]$, $L_{\star}$) can be combined with stellar
evolution tracks to determine the stellar mass, radius, density,
etc. (e.g., Ford et al.\ 1999), with some well-known degeneracies
(e.g., hook region).  For relatively bright and nearby stars in the
California and Carnegie Planet Search and Hipparcos catalog, this
method has been used to estimate stellar parameters such as the mass
and radius (Valenti \& Fischer 2005; Takeda et al.\ 2007).  We applied
the Bayesian stellar parameter estimation code of Takeda et al.\ (2007) to
calculate density directly from the joint posterior distribution,
accounting for correlations between parameters and the non-linear
transformation between observable and derived quantities.  We find
that this method can determine the mean stellar density with an
accuracy of $\simeq 4-10\%$, with random uncertainties dominated by
the parallax.

We caution that there may also be systematic uncertainties due to the
stellar evolutionary tracks.  Such systematic effects could lead to
$\tau_o$ or $\tau_b$ being systematically over- or under-estimated and
that the error could be highly correlated with the stellar type or
other stellar parameters.  Ideally, such systematics would be
mitigated if the densities of stellar models were validated by
independent observations such as double-lined eclipsing binaries for
similar type stars.  Both the CoRoT and Kepler missions will
contribute to stellar astronomy and are likely to contribute towards
improving and testing the precision of stellar models.  In the absense
of externally validated models, systematic uncertainties in stellar
models may limit this techinque for systems with small eccentricities.
Fortunately, the stellar density enters only to the one third power,
so potential systematic effects are unlikely to be significant for
most planets that have sizable eccentricities.  In practice, the
eccentricity analysis should be coupled to a sensitivity analysis that
tests whether conclusions are sensitive to the choice of stellar models.

Unfortunately, many target stars for Kepler will be too faint and
distant to have well determined parallaxes.  One alternative approach
is to replace $L_{\star}$ with a spectroscopically determined stellar
surface gravity ($\log g$).  Unfortunately, $\log g$ is quite
difficult to measure precisely.  For a high-quality spectroscopic
observation (e.g., $\sigma_{T_{\rm eff}}\sim 100$K and $\sigma_{\rm
[Fe/H]}\sim 0.1$dex), the formal uncertainty in $\log g$ would
typically be $\sim 0.1$dex.  Unfortunately, there are typically
significant correlations with both $T_{\rm eff}$ and $[Fe/H]$ (Valenti
\& Fischer 2005).  For a main sequence solar-type star, the uncertainties in 
the atmophseric parameters would
translate to an uncertainty in $\rho_\star^{1/3}$ (and hence $\tau_b$)
of $\simeq 8$\%.  In some cases, additional constraints on stellar
properties such as rotation rate, strength of Ca HK emission, presence
of Li, and/or asteroseismology may be able to further constrain
stellar parameters and improve the stellar sensitivity for measuring
eccentricities.

\section{Statistical Methodology}
\label{SecStats}
In this section, we outline a few of the statistical approaches that
could be applied to translate measurements of $\tau_o$ and/or
$\tau_b$ into eccentricity constraints.  First, we outline a Bayesian
approach to calculating the joint posterior probability distribution
for the eccentricity and argument of periastron for individual
planets.  These constraints would be particularly valuable when combined with
additional constraints on dynamical properties of the system.
Second, we show that the distribution of normalized transit durations
contains significant information about the eccentricity distribution
of a population of planets.  

\subsection{Individual Planets}
\label{SecStatsMcmc}
\begin{figure}
\plotone{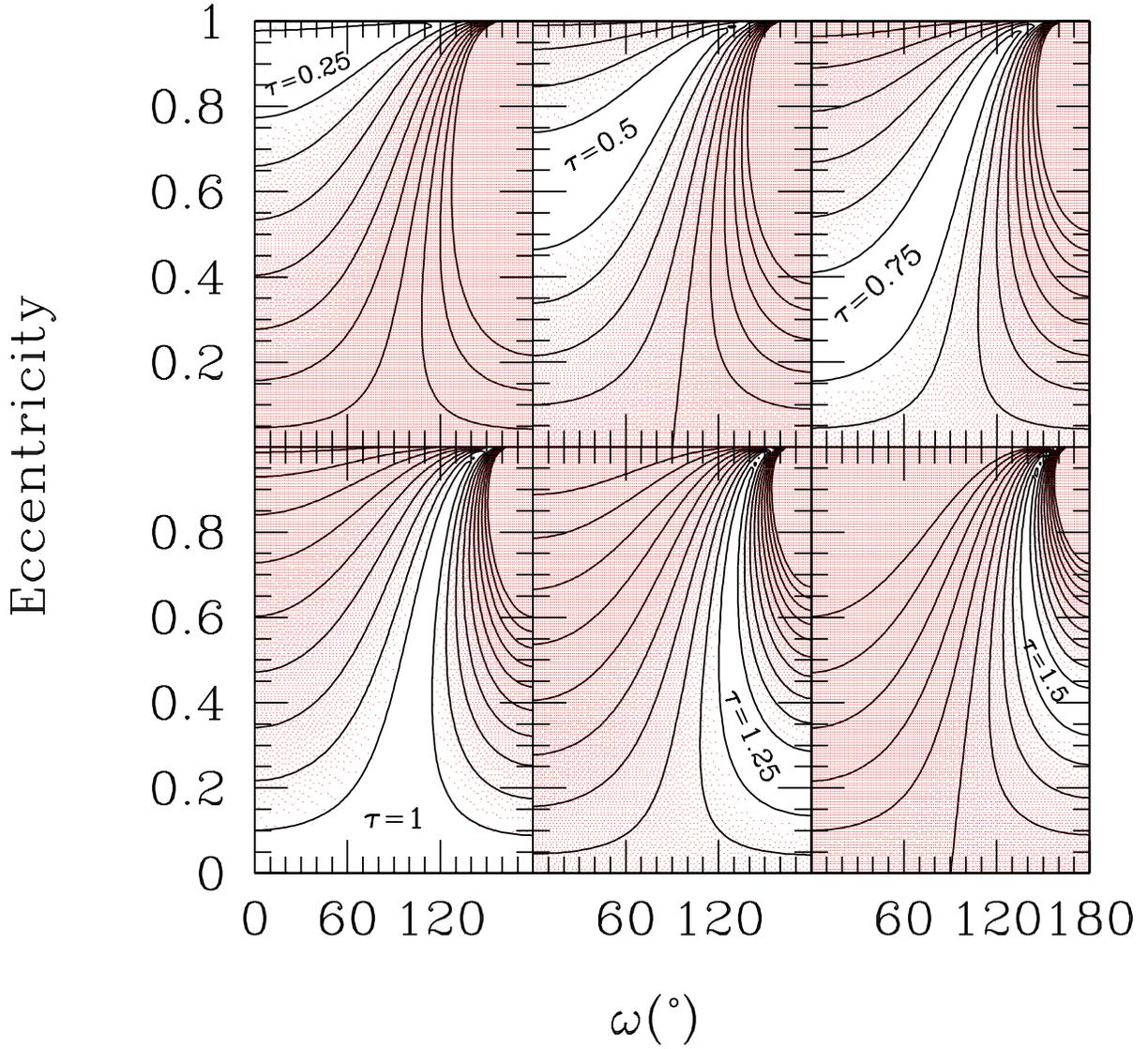} 
\caption{
Posterior Joint Probability Distribution for $e$ and $\omega$:
 Here we consider the eccentricity constraint based
on the measured value of $\tau_b$ for an individual planet with a
measured impact parameter.  We show contours equivalent to 1, 2,
3,...-$\sigma$ bounds on the combination of eccentricity and argument
of pericenter.  The shaded regions are excluded by the measurement of
$\tau_b$.  The panels correspond to $\tau_b=0.25$ (top left), 0.5
(top center), 0.75 (top right), 1 (bottom left), 1.25 (bottom center),
and 1.5 (bottom right).  Here we assume that the detection probability
(given that a transit occurs) is independent of transit duration and
that the measurement of $\hat{\tau}_b$ is normally distributed and has
a standard deviation, $\sigma_{\tau_b}=0.1$.
\label{FigTauContours} }
\end{figure}

For planets with high signal-to noise transit light curves and
well-measured stellar properties, we suggest a Bayesian framework for
determining the constraints on the eccentricity and argument of
pericenter.  To illustrate this method, we adopt non-informative
prior probability distributions of $p(e)=1$ for $0\le e<1$, $p(\omega)=1/(2\pi)$ for $0\le \omega < 1$, and
$p(\cos i)=1/2$ for $-1 \le \cos i < 1$.  We assume that the observed $\hat{\tau}_b$ is
normally distributed about the true value of $\tau_b$ (i.e.,
$\hat{\tau}_b \sim N(\tau_b,\sigma^2_{\tau_b}))$ with
$\sigma_{\tau_b}$ reflecting the uncertainty in the stellar parameters.
In Fig.\ \ref{FigTauContours}, we show examples of the joint posterior
probability distribution for six possible values of $\hat{\tau}_b$ .
Note that in this figure, $\omega$ is measured relative to the line of sight,
not relative to the plane of the sky, as is typical when using radial velocity observations.
For actual systems, Markov chain Monte Carlo based simulation methods
(Ford 2005, 2006) could be used to calculate posterior probability
distributions allowing for parameters with correlated uncertainties
and/or non-trivial error distributions (e.g., Holman et al.\ 2006;
Burke et al.\ 2007; Takeda et al.\ 2007), as well as eccentric orbits.

\subsection{Characterizing the Eccentricity Distribution of Transiting Planets}
\label{SecStatsKs}
\begin{figure}
\plotone{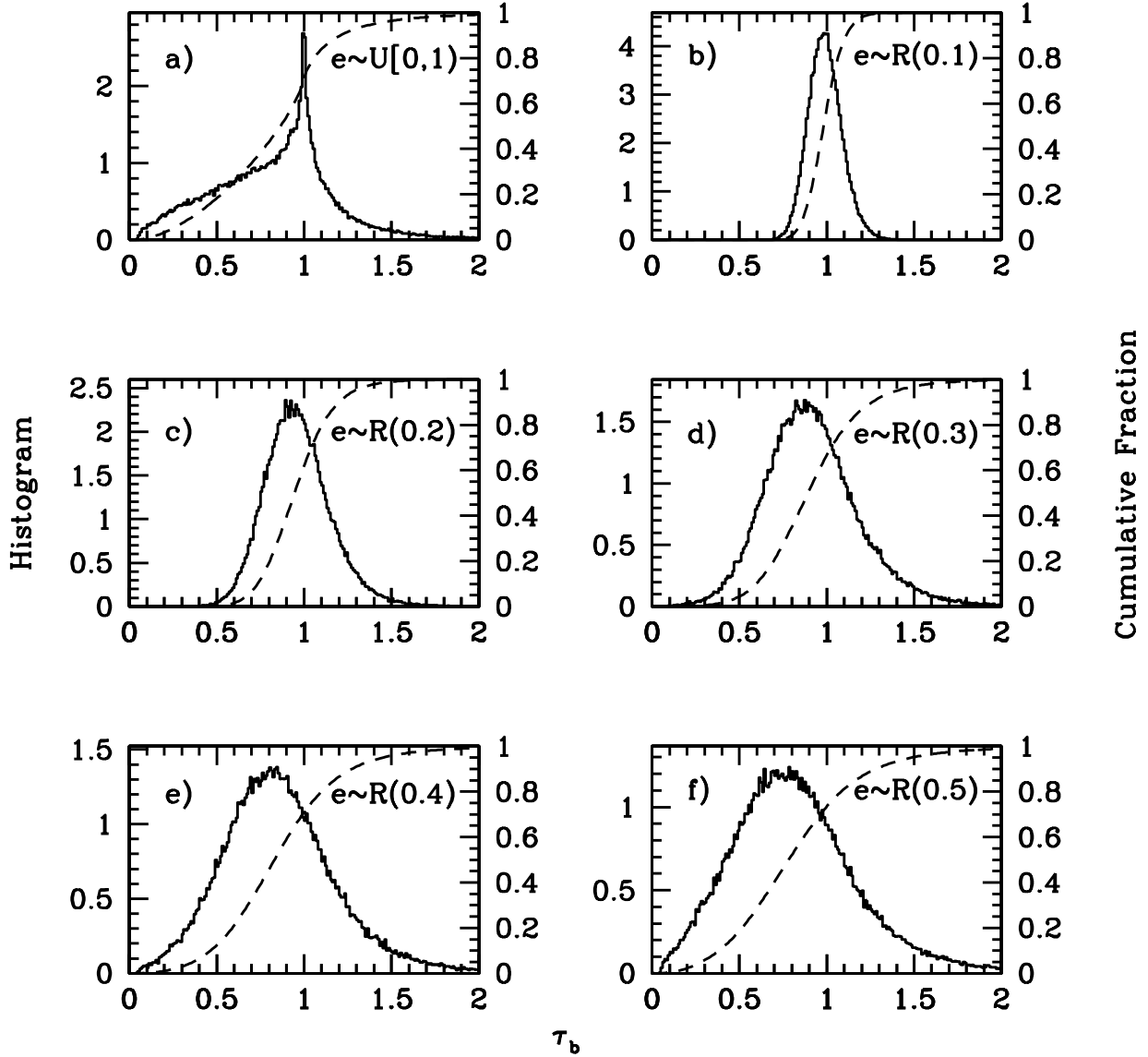} 
\caption{Distribution of Transit Duration for Various Eccentricity Distributions:
We show histograms (solid curves) and cumlative distributions (dashed curves) of $\tau_b$, the ratio of the observed transit duration to
that expected for a similar planet, star, impact parameter, and
orbital period, but a circular orbit.  Panel a corresponds to a
uniform eccentricity distribution, panels b-f correspond to a Rayleigh
eccentricity distribution with Rayleigh parameter of 0.1 (b), 0.2 (c),
and 0.3 (d), 0.4 (e), and 0.5 (f).
\label{fig_hist_wb}
}
\end{figure}

We have performed Monte Carlo simulations to calculate the expected
distribution of transit durations for various eccentricity
distributions.  Here, we consider two limiting cases corresponding to
low and high signal-to-noise observations: a) only the transit depth,
duration, and orbital period are measured, so that the observed light
curve provides no information about the impact parameter (Fig.\
\ref{fig_hist_wob}), and b) the time of all four points of contact are
measured so that the observed light curve provides a good estimate of
the impact parameter (Fig.\ \ref{fig_hist_wb}).

\subsubsection{Impact Parameter Not Measured}
\label{SecStatsKsLow}
While it would be most desirable to measure the eccentricity for each
individual planet, this will not always be possible.  For some planets
(particularly those around faint stars), the light curve will not be
measured with sufficient precision to measure the orbital inclination
and sufficient high-precision RV observations will be impractical.  In
these cases, it will still be possible to constrain the mean
eccentricity of a population of such planets.  Since these challenges
will be most common for faint stars, we expect that missions such as
CoRoT and Kepler will detect a large number of such planets available
for statistical analyses.

\begin{figure}
\plotone{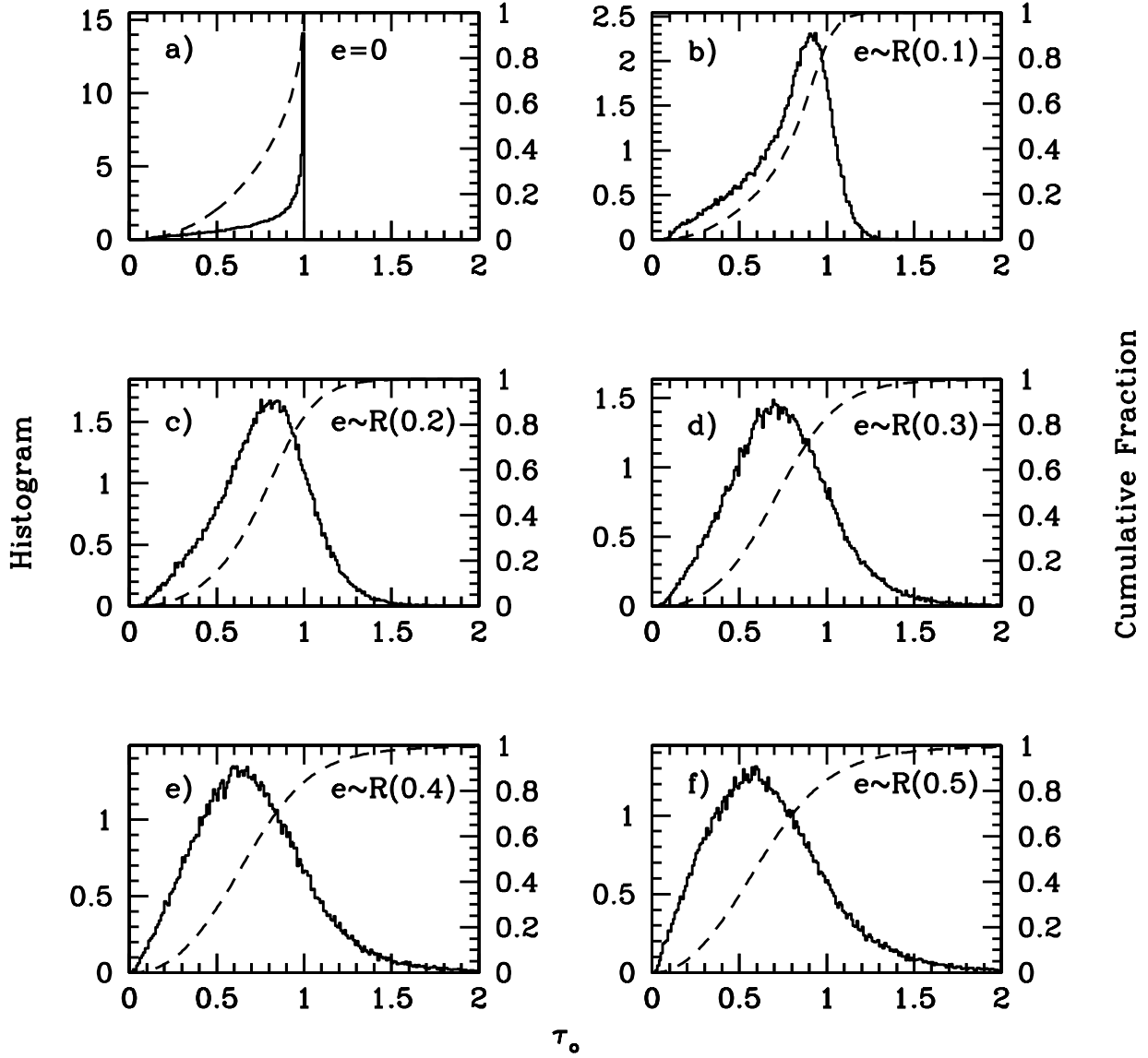} 
\caption{Distribution of Transit Duration for Various Eccentricity Distributions:
We show histograms (solid curves) and cumlative distributions (dashed curves) of $\tau_o$, the ratio of the observed transit duration to
that expected for the same planet, star, and orbital period, but a
circular orbit and a central transit.  Panel a corresponds to only
circular orbits, panels b-f correspond to a Rayleigh eccentricity
distribution with Rayleigh parameter of 0.1 (b), 0.2 (c), and 0.3 (d),
0.4 (e), and 0.5 (f).
\label{fig_hist_wob}  
}
\end{figure}

\begin{figure}
\plotone{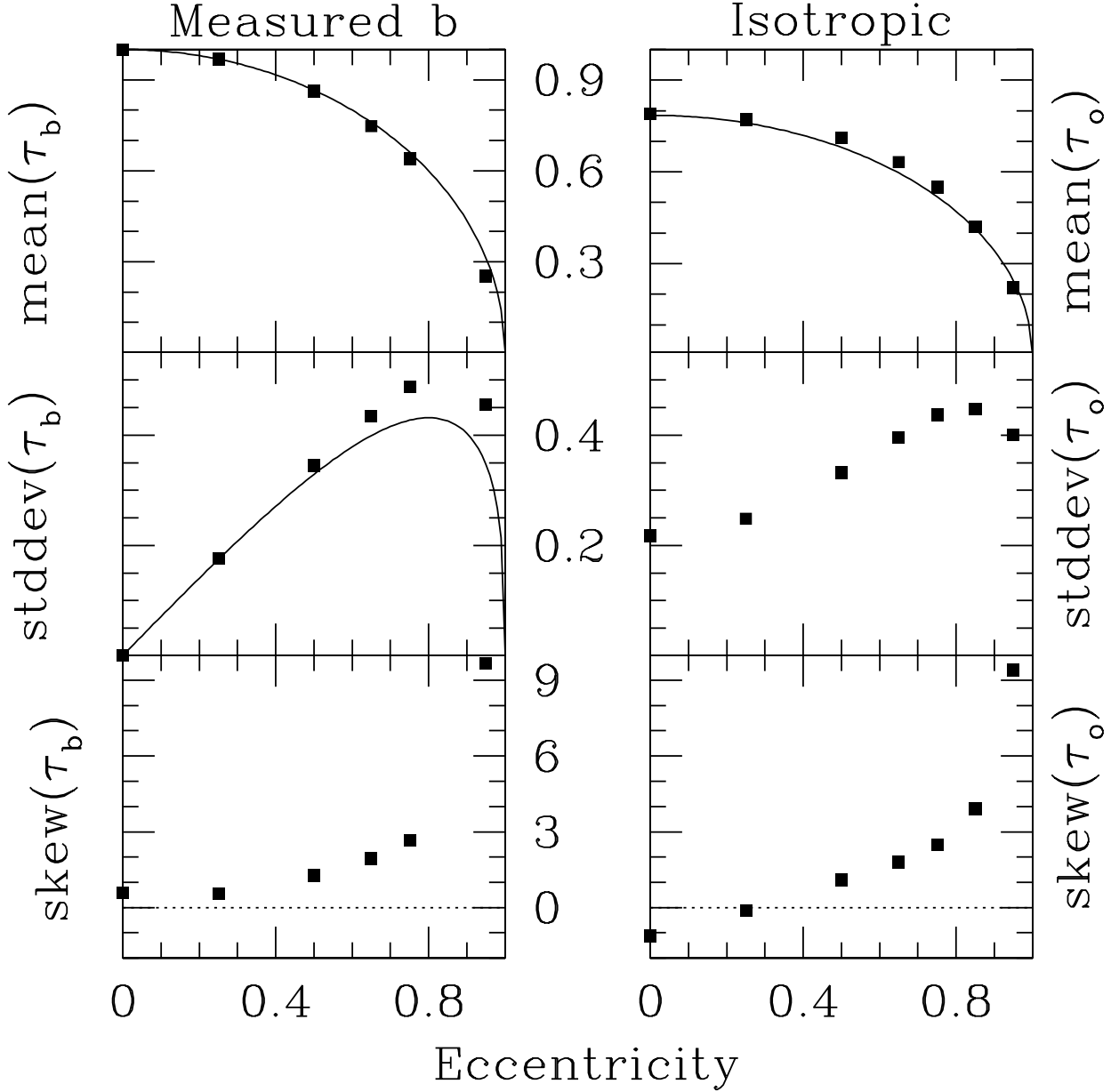} 
\caption{Moments of the Normalized Transit Duration Distribution:
 Here we consider several ensembles of transiting
planets, each with a single fixed eccentricity.  For each
eccentricity, we calculate the distribution of $\tau_b$ (left; for
transits with measured impact parameters) and $\tau_o$ (right; for
transits without measured impact parameters).  Here we plot the mean,
standard deviation, and skewness for both both $\tau_b$ and $\tau_o$
for each eccentricity.  We show analytic approximations with curves
when available.
\label{fig_moments} }
\end{figure}

We consider the distribution of $\tau_o\equiv t_{D} / t_{D,o}$, where
$t_D$ is the actual transit duration, $t_{D,o}$ is the transit
duration expected for the same planet and star, but assuming a
circular orbit ($e=0$) and central transit ($b=0$).  The distribution
for the impact parameter, $b$, is determined by assuming that the
inclination is distributed isotropically (subject to the constraint
that a transit occurs). We also assume that the observed
$\hat{\tau}_o$ is normally distributed about the true value of
$\tau_o$ (i.e., $\hat{\tau}_o \sim N(\tau_o,\sigma^2_{\tau_o}))$ with
$\sigma_{\tau_o}\simeq0.05-0.15$ due to uncertainty in the stellar
parameters (depending on the stellar properties and available
follow-up observations).
In Fig.\ \ref{fig_hist_wob}, we show the distribution of $\tau_o$ for
six eccentricity distributions.  
In Fig.\ \ref{fig_moments} (right), 
we plot the mean, standard deviation,
and skewness of $\tau_o$ for an ensemble of planets with a
given eccentricity.
For planets with a circular orbit and an isotropic distribution of
inclinations, $\tau_o$ will have a mean of $\simeq0.79$ and a standard
deviation of $\simeq0.22$ (Fig.\ \ref{fig_hist_wob}, panel a).
If we assume a uniform distribution of eccentricities ($e\sim U[0,1)$), then the mean $\tau_o$ decreases to 0.64 and the standard
deviation increases to 5.2.
Alternatively, if we instead assume an eccentricity distribution
similar to that observed for giant planets ($e\sim R(0.3)$, a Rayleigh
distribution with Rayleigh parameter 0.3; Juric \& Tremaine 2007),
then the mean decreases to $\simeq0.74$ and the standard deviation
increases to $\simeq0.29$.
Therefore, it would be possible to distinguish between all three
models with a plausible sample of planets, assuming normally
distributed errors and asymptotic scalings for the mean of the
distribution, even without measuring any impact parameters or RVs.
If we were to use only the mean value of $\tau_o$, then we
would be ignoring the variances and shape of the distribution.
Instead, a Komogorov-Smirnov test can be used to compare the observed
distribution of $\tau_o$ to the $\tau_o$ distribution predicted by a
given model.  

To determine how large an observed sample is needed to obtain
statistically significant results, we perform Monte Carlo simulations.
For each trial, we generate two samples of transiting planets.  The
first sample represents a hypothetical observed sample of $N_{pl}$
planets.  The second sample is much larger and is used to calculate
the predicted distribution of $\tau_o$ for a given theoretical model.
For both samples, we assume an isotropic distribution of viewing
angles, i.e., $\cos i \sim U[-1,1)$ and $\omega \sim U[0,2\pi)$. For
each trial, we test whether a Kolmogorov-Smirnov test can reject the
null hypothesis that the two distributions of $\tau_o$ are drawn from
a common distribution.  For each $N_{pl}$ we perform several thousand
trials; we increase $N_{pl}$ until we find that the null hypothesis is
rejected at the $95\%$ confidence level for at least half of the
trials.  We present our results in Table \ref{tab_one}.  Our
simulations show that a population of planets on circular orbits ($e =
0$) could be distinguished from either $e\sim U[0,1)$ (eccentricities
distributed uniformly between zero and unity) or $e\sim R(0.3)$ (a
Rayleigh distribution with Rayleigh parameter of 0.3) by using
measurements of $\tau_o$ for $N_{pl}\go30$ planets, even without
measuring any impact parameters or RVs.  Distinguishing an
observed population with eccentricities distributed as $e\sim U[0,1)$
from a model with $e\sim R(0.3)$, would require $N_{pl}\go 117$ planets.
Distinguishing between an observed population with eccentricities
distributed as $e\sim R(0.3)$ rather than a model with $e\sim U[0,1)$, would
require $N_{pl}\go 174$ planets.  The difference is due to the fact
that we use a two-sample Kolmogorov-Smirnov test and the sample size
for the theoretical model is at least an order of magnitude greater
than the sample size for the sample of ``observed'' planets.

\subsubsection{Impact Parameter Measured}
\label{SecStatsKsHi}
\begin{figure}
\plotone{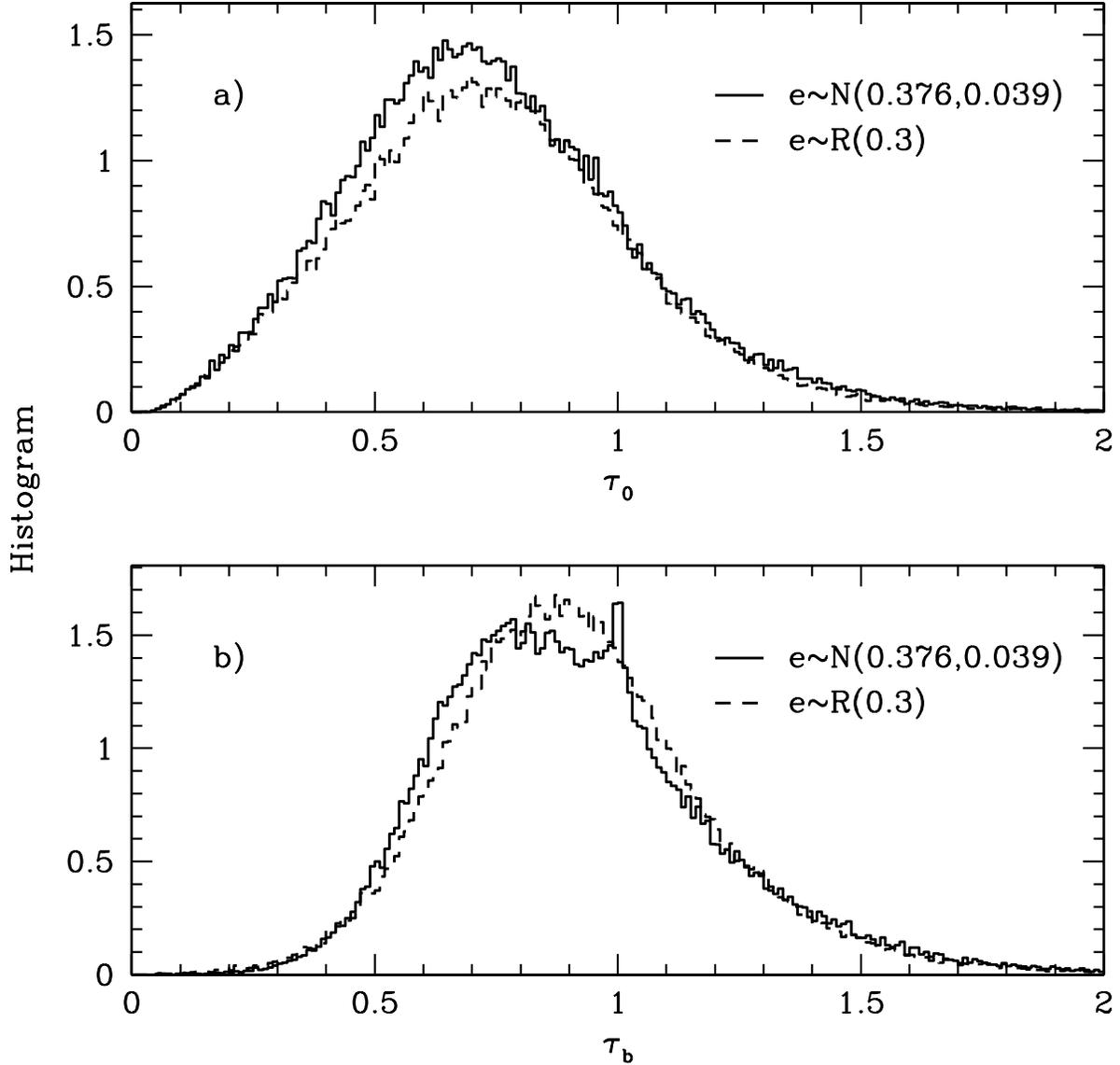} 
\caption{Distribution of Transit Duration for Eccentricity Distributions with Common Mean and Variance:
{\em Top:} Histograms of $\tau_o$, the ratio of the observed transit
duration to that expected for the same planet, star, and orbital
period, but a circular orbit and a central transit.  We show results
for two eccentricity distributions: normal (solid) and Rayleigh
(dashed).  The mean and variance of the normal distribution have been
chosen to match that of a Rayleigh distribution with Rayleigh
parameter 0.3.  The resulting distributions of $\tau_o$ are so similar that it would be extremely difficult to distinguish between these eccentricity distributions.
{\em Bottom:} Same as above, but for histograms of $\tau_b$, the ratio
of the observed transit duration to that expected for a similar
planet, star, impact parameter, and orbital period, but a circular
orbit.  Again, the distributions of $\tau_b$ are too similar to distinguish between the eccentricity distributions.
\label{FigDistribComp}
}
\end{figure}

When high quality photometry measures the impact parameter ($b$), then
we can analyze the distribution of $\tau_b\equiv t_{D} / t_{D,b}$,
where $t_{D,b}$ is the transit duration expected for the same planet
with the same impact parameter, but assuming a circular orbit.  
If all transiting planets had a single eccentricity, then the expected
mean value of $\tau_b$ is given by $\left<\tau_b\right>\simeq
\sqrt{1-e^2}$ (Tingley \& Sackett 2005) and the expected standard
deviation is $\left<\sigma_{\tau_b}\right> \simeq
(1-e^2)^{3/8}\sqrt{1-(1-e^2)^{1/4}}$.
We show the distribution of $\tau_b$ for several eccentricity
distributions in Fig.\ \ref{fig_hist_wb} and plot the mean, standard
deviation, and skewness for sample distributions of $\tau_b$ in Fig.\
\ref{fig_moments} (left).  Note that for small eccentricities, the
distribution of $\tau_b$ is strongly peaked, making it possible to
perform a significant test of the null hypothesis that terrestrial
planets have nearly circular orbits with a small sample size
(\S\ref{SecStatsNull}).  To account for the uncertainty in the stellar
parameters, we assume $\sigma_{\hat{\tau}_b}\simeq 0.05-0.15$ and
normally distributed errors and asymptotic scalings.  Then, it would
be possible to distinguish between various models for the eccentricity
distribution of observed planets with a few tens of transiting planets
(see Table \ref{tab_two}).  We find that the distribution of $\tau_b$
provides significant constraints on both the mean and the width of the
eccentricity distribution.  For example, Monte Carlo simulations using
the Kolmogorov-Smirnov test (similar to those described in
\S\ref{SecStatsKsLow}, but replacing $\tau_o$ with $\tau_b$) show that
distinguishing between two Gaussian eccentricity distributions with
mean eccentricities of 0.18 and 0.36 would require $N_{pl}\go30$
planets.  Similarly, distinguishing between two Gaussian eccentricity
distributions with a common mean of 0.38 and standard deviations of
0.10 and 0.24 would require $N_{pl}\go 65$ planets based on a
Kolmogorov-Smirnov test and a 95\% confidence level.  However, it will
be very difficult to measure the higher order moments of the
eccentricity distribution (see Fig.\ \ref{FigDistribComp}).  In Table
\ref{tab_two}, we list the number of planet required for several
additional pairs of eccentricity distributions.

\subsection{Testing the Null Hypothesis of Circular Orbits for Individual Planets}
\label{SecStatsNull}
We are particularly interested in addressing the question, ``How
frequently are terrestrial planets on nearly circular orbits?''
Therefore, we suggest two statistical tests of the null-hypothesis
that each planet is on a circular orbit.  

\subsubsection{Low S/N Light Curves \& Long Duration Transits}
For an eccentric orbit with the transit near apocenter, the transit
duration can be significantly greater than would be expected for the
observed orbital period.  It is possible to measure a minimum
eccentricity for long transits, even for low S/N transits and without
measuring RV parameters.  Since a non-zero impact parameter can only
decrease the duration of transit (relative to a central transit), it
will be possible to reject the null hypothesis of a circular orbit
(and hence detect a non-zero eccentricity) for planets with long
duration transits such that $\tau_o$ or $\tau_b$ greater than unity.
Unfortunately, the geometric probability of a transit occurring is
greater for short transits when the planet is closer to the star.
Therefore, if eccentric planets are common, then there will be more
transits with $\tau<1$.  Fortunately, even a small number of long
transits can provide significant constraints for the eccentricity
distribution.

\subsubsection{High S/N Light Curves \& Stellar Models}

For short-period planets that can be assumed to have
tidally circularized, $\tau_b =1$, so high precision light curve
observations may provide additional constraints on the star's
properties.  For example, Holman et al.\ (2007) measure
$\rho_{\star}^{1/3}$ to $\simeq 1.5$\% accuracy for TrES-2 (V=11.4).
When combined with theoretical models, this allows for very accurate
determinations of the stellar and planet radii (Sozzetti et al.\
2007). 
This will enable tests of the null-hypothesis that a
planet is on a nearly circular orbit.  If it is accurate, then the
stellar properties would be measured quite precisely and all
measurements should be self-consistent.  On the other hand, if the
 planet is actually eccentric, then assuming a circular orbit could
result in an inconsistency.
For example, by comparing the allowed stellar models to a
spectroscopic measurement of $\log g$ one could recognize planets with
eccentricities exceeding $\simeq0.15$ for solar-like stars.
Alternatively, the putative location of the star in the ($T_{\rm
eff}$, $\rho_{\star}$, [Fe/H]) parameter space could be inconsistent
with any stellar model (e.g., Sozzetti et al.\ 2007).  The sensitivity
of this test will vary significantly with $T_{\rm eff}$ and be most
powerful for stars slightly cooler than the Sun.  Unfortunately, this
method is unlikely to be effective for planets with orbital periods
larger than a few days, since they may have eccentric orbits.

\subsection{Role of Radial Velocity Observations}
\label{SecRV}
The realization that many giant planets are on eccentric orbits was
the result of RV surveys (e.g., Butler et al.\ 2006 and references
therein).  Clearly, RV observations can measure orbital
eccentricities, provided that there is a sufficient number, phase coverage, and time
span of observations with sufficiently high precision (Ford 2005).  In
fact, the observational constraints from radial velocity observations
and the photometric method that we describe are complimentary.  Our
photometric method is most sensitive for measuring eccentricities of
planets if the pericenter direction is pointing towards or away from
the observer (Fig.\ \ref{FigTauContours}).  On the other hand, the
radial velocity method is most sensitive to measuring eccentricities
of planets with pericenter nearly in the plane of the sky (Ford 2005;
Laughlin et al.\ 2005).

Unfortunately, many of the CoRot and Kepler target stars (V$\sim9-16$)
will be much fainter than the typical targets of RV surveys
(V$\sim5-9$).  Ground-based transit surveys have obtained follow-up RV
observations for candidate planets, but typically at a relatively
modest precision.
While there are plans to obtain RV confirmation of planet candidates
identified by CoRoT or Kepler, these observations will require a
large investment of telescope time, especially when attempting to
detect Earth-mass or even Neptune-mass planets around relatively faint
stars.
Even when RV observations can confirm terrestrial planet candidates
found by CoRoT or Kepler, they will likely make use of the known orbital
period and phase to observe at near the extrema of the RV
curve.  Measuring the eccentricity requires measuring the shape of the
curve and hence many observations spread across a broad range of orbital
phases. 
The amplitude of the deviations of the RV curve from that expected for
the same planet on a circular orbit is less than the total RV
amplitude by a factor of $e$ (to lowest order in
eccentricity). Assuming a single planet with a orbital period known
from photometry, a detection of eccentricity for a planet with
$e\simeq 0.3$ will typically require an order of magnitude more
observing time than would be required to detect the planet's total RV
amplitude (at the same level of significance and assuming observations
nearly evenly distributed in orbital phase).  As an additional
complication, the RV method measures the reflex velocity of the star
induced by all planets, not just one transiting planet.  If a star
harbors a (potentially undetected) non-transiting planet with a RV
amplitude comparable to or greater than the radial velocity amplitude due
to the epicyclic motion of the transiting planet, then RV observations
would face an even greater challenge in measuring the transiting
planet's eccentricity.
Despite these challenges, we certainly encourage radial velocity
observations for constraining eccentricities whenever they are practical.

\section{Discussion}
\label{SecDiscuss}
We have described how photometric observations can constrain the
eccentricities of individual transiting planets and characterize the
eccentricity distribution of a population of planets.  For each
planet, we can test the null hypothesis that it is on a circular orbit
and calculate the posterior probability distribution for the
eccentricity.  A combination of such analyses for several transiting planets
could be used to characterize the eccentricity distribution of a
population of planets.  For example, this method could be used to
investigate how the fraction of eccentric orbits varies with the
orbital period or physical proprieties of the star and planet.  
We expect this type of analysis to become increasingly powerful given
the rapidly growing number of known transiting planets. 

This type of analysis will be particularly valuable for low-mass
transiting planets or transiting planets around faint host stars.
In both cases, radial velocity follow-up observations will be
extremely challenging.  This will be the case for many transiting
planet candidates found by space-based transit searches, such as CoRoT
and Kepler.  The capability of these missions to discover
terrestrial-mass planets is particularly exciting.  Our method could
could determine if terrestrial planets with low eccentricities like
the Earth are common or rare.  This would provide significant constraints
on theories proposed to explain the eccentricities of extrasolar planets.
For example, Kepler
might find many terrestrial planets with low eccentricity orbits,
suggesting that the mechanisms that excite the eccentricities of giant
planets are often ineffective for terrestrial mass planets.  In this
scenario, those terrestrial planets that do have large eccentricities might
typically be accompanied by nearby giant planets, suggesting that it
is the giant planets are responsible for exciting the eccentricities
of terrestrial planets (Veras \& Armitage 2005, 2006).  Alternatively,
Kepler might find that terrestrial planets are much more common than
giant planets, and yet they still commonly have large eccentricities.
This could arise from eccentricity excitation mechanisms that do not
require giant planets, or due to interactions with previous giant
planets that have since been ejected, accreted or destroyed by the star (e.g.,
Ford et al.\ 2005; Raymond et al.\ 2006; Mandell et al.\ 2007).

Our method also provides a means for studying the tidal evolution of
short-period planets.  Tidal effects are likely to circularize planets with sufficiently
short orbital periods.  For short-period planets, we can compute the tidal
circularization timescale ($t_{\rm circ}$) based on the properties of
the star and planet.  If there is a sharp transition between circular
and eccentric orbits, then this could be used to place constraints on
tidal theory (e.g., Zahn \& Bouchet 1989; Melo et al.\ 2001; Mathieu
et al.\ 2004).  For eccentric planets with relatively short $t_{\rm
circ}$, it may be possible to place a lower limit on the $Q$ factor
the is related to the planet's internal structure (e.g., Ford et al.\
1999; Bodenheimber et al.\ 2001; Maness et al.\ 2007; Mardling 2007).

In a Bayesian framework, one can calculate the
posterior probability distribution for the fraction of each orbit
during which the planet-star separation is between an inner and outer
cut-off.  If the cut-offs are set to be the putative boundary of the
habitable zone, then we can then ask, ``What fraction of terrestrial
planets are in the habitable zone for some/at least half/all of their
orbit?''.  The results of such investigations could have implications
for the climates of potentially habitable planets, the frequency of
such planets, and the design of future missions that aim to detect and
characterize nearby Earth-like planets that could harbor life (e.g.,
Marcy et al.\ 2005).

We have demonstrated that it is practical to collect sufficient
photons to characterize the eccentricity distribution of terrestrial
extrasolar planets, but we assumed that limb-darkening parameters can
be well constrained by some combination of stellar modeling and
external observations.  Our analytical estimates have not incorporated
limb darkening effects or potential systematic uncertainties in
stellar models.  Future search should address both of these effects.
Multi-wavelength observations (particularly in the infrared) could be
particularly useful for addressing both these issues.  In particular,
we plan to investigate the potential for combinations of space-based
detections and ground-based follow-up observations to improve the
characterization of the eccentricities of transiting planets.

Finally, we note that this method for characterizing the
eccentricities of terrestrial planets from transit light curves
underscores the importance of developing and validating precise and
accurate stellar models.  Uncertainties in stellar parameters models
are expected to dominate the error budget for bright target stars.
The potential for systematic uncertainties due to stellar modeling
will make it particularly challenging to study low eccentricity
systems as a function of stellar properties.  Fortunately, we these
concerns would not preclude our method from being applied to
terrestrial-sized planets recognizing the relatively large
eccentricities typical for giant planets.

\acknowledgements We thank G. Bakos, D.\ Charboneau, M.\ Holman, D.\
Latham, G.\ Takeda, and an anonymous referee for discussions and
suggestions.
Support for E.B.F.\ was provided by NASA through Hubble Fellowship
grant HST-HF-01195.01A awarded by the Space Telescope Science
Institute, which is operated by the Association of Universities for
Research in Astronomy, Inc., for NASA, under contract NAS 5-26555.
Additional support for E.B.F.\ and D.V.\ was provided by the
University of Florida.

\clearpage
              
\begin{deluxetable}{ccccccccc}
\rotate
\tabletypesize{\small}
\tablewidth{0pt}
\tablecaption{Number of Planets Required to Distinguish between Distributions (Low S/N)\label{tab_one}}
\tablehead{
\colhead{ } & \colhead{$e=0$} & \colhead{$e\sim R(.1)$} & \colhead{$e\sim R(.2)$} & \colhead{$e\sim R(.3)$} & \colhead{$e\sim R(.4)$} & \colhead{$e\sim R(.5)$} & \colhead{$e\sim R(.6)$} & \colhead{$e\sim U[0,1)$}}
\startdata
$e=0$          & \nodata & 98 & 44 & 30 & 22 & 19 & 16 & 26 \\
$e\sim R(.1)$  & 78 & \nodata & 187 & 53 & 33 & 23 & 20 & 34 \\
$e\sim R(.2)$  & 51 & 171 & \nodata & 250 & 84 & 42 & 33 & 52 \\
$e\sim R(.3)$  & 36 & 54 & 180 & \nodata & 355 & 117 & 67 & 174 \\
$e\sim R(.4)$  & 25 & 29 & 68 & 270 & \nodata & 575 & 215 & 675 \\
$e\sim R(.5)$  & 20 & 24 & 42 & 108 & 540 & \nodata & $>$1000 & 975 \\
$e\sim R(.6)$  & 17 & 19 & 28 & 55 & 155 & 850 & \nodata & 255 \\
$e\sim U[0,1)$ & 26 & 29 & 53 & 117 & 540 & 850 & 310 & \nodata \\
\enddata
\tablenotetext{a}{We list the number of planets required to 
distinguish an observed distribution (taken from the top row) from a
theoretical distribution (taken from the left column) at the 95\%
confidence level based on a Kolmogorov-Smirnov test applied to
$\tau_o$.}
\end{deluxetable}

\begin{deluxetable}{ccccccccc}
\rotate
\tabletypesize{\footnotesize}
\tablewidth{0pt}
\tablecaption{Number of Planets Required to Distinguish between Distributions (High S/N)\label{tab_two}}
\tablehead{
\colhead{ } & \colhead{$e=0$} & \colhead{$e\sim R(.1)$} & \colhead{$e\sim R(.2)$} & \colhead{$e\sim R(.3)$} & \colhead{$e\sim R(.4)$} & \colhead{$e\sim R(.5)$} & \colhead{$e\sim R(.6)$} & \colhead{$e\sim U[0,1)$}}
\startdata
\cutinhead{$\sigma_{\tau_b} = 0.05$}
$e=0$          & \nodata & 14 & 11 & 8 & 7 & 7 & 6 & 8 \\
$e\sim R(.1)$  & 17 & \nodata & 47 & 20 & 12 & 9 & 9 & 16 \\
$e\sim R(.2)$  & 11 & 45 & \nodata & 77 & 33 & 18 & 17 & 30 \\
$e\sim R(.3)$  & 8 & 15 & 72 & \nodata & 158 & 58 & 34 & 108 \\
$e\sim R(.4)$  & 7 & 9 & 21 & 122 & \nodata & 380 & 127 & 385 \\
$e\sim R(.5)$  & 6 & 7 & 14 & 46 & 335 & \nodata & 725 & 275 \\
$e\sim R(.6)$  & 5 & 6 & 11 & 28 & 99 & 495 & \nodata & 100 \\
$e\sim U[0,1)$ & 8 & 11 & 20 & 67 & 355 & 305 & 133 & \nodata \\
\cutinhead{$\sigma_{\tau_b} = 0.15$}
$e=0$          & \nodata & 29 & 19 & 13 & 10 & 9 & 8 & 11 \\
$e\sim R(.1)$  & 35 & \nodata & 73 & 23 & 17 & 11 & 12 & 22 \\
$e\sim R(.2)$  & 19 & 65 & \nodata & 88 & 43 & 23 & 22 & 38 \\
$e\sim R(.3)$  & 13 & 23 & 97 & \nodata & 210 & 70 & 47 & 125 \\
$e\sim R(.4)$  & 8 & 14 & 31 & 145 & \nodata & 465 & 148 & 480 \\
$e\sim R(.5)$  & 8 & 11 & 18 & 60 & 355 & \nodata & 775 & 375 \\
$e\sim R(.6)$  & 7 & 8 & 13 & 37 & 120 & 545 & \nodata & 135 \\
$e\sim U[0,1)$ & 10 & 14 & 26 & 75 & 410 & 475 & 175 & \nodata \\
\enddata
\tablenotetext{a}{We list the number of planets required to
distinguish an observed distribution (taken from the top row) from a
theoretical distribution (taken from the left column) at the 95\%
confidence level based on a Kolmogorov-Smirnov test applied to
$\hat{\tau}_b$ with $\sigma_{\tau_b} = 0.15$ (bottom) and 0.05 (top).}
\end{deluxetable}

\clearpage


%

%
%
%
%
%
%
%

%
%

\end{document}